\documentclass[prl,showpacs,floatfix,twocolumn]{revtex4}
\usepackage{graphicx}
\newcommand{\be}{\begin{eqnarray}}
\newcommand{\ee}{\end{eqnarray}}
\def\beq{\begin{equation}}
\def\eeq{\end{equation}}

\begin{document}
\title{The Anderson transition in quantum chaos}
\author{Antonio M. Garc\'{\i}a-Garc\'{\i}a}
\affiliation{Physics Department, Princeton University, Princeton, New Jersey 08544, USA}
\author{Jiao Wang}
\affiliation{Temasek Laboratories, National University of Singapore,119260 Singapore}
\begin{abstract}
We investigate the effect of classical singularities in the quantum properties
of non-random Hamiltonians. We present explicit results for the case of a kicked
rotator with a non-analytical potential though extensions to higher 
dimensionality or conservative systems are straightforward. 
It is shown that classical singularities produce anomalous diffusion in the 
classical phase space. Quantum mechanically, the eigenstates of the evolution 
operator are power-law localized with an exponent universally given by the 
type of classical singularity.
For logarithmic singularities, the classical motion presents $1/f$ noise and
the quantum properties resemble those of an Anderson transition, namely, 
multifractal eigenstates and critical statistics. Neither the classical nor
the quantum properties depend on the details of the potential but only on the
type of singularity. It is thus possible to define new universality class in 
quantum chaos by the relation between classical singularities (anomalous 
diffusion) and quantum power-law localization. 
\end{abstract}
\maketitle

The description of quantum features based on the underlying classical dynamics 
has been intensively investigated since the early days of quantum mechanics.
For instance, the celebrated Bohigas-Giannoni-Schmit conjecture \cite{oriol} 
states that the spectral correlations of a quantum system whose classical 
counterpart is fully chaotic depend only on the global symmetries of the 
system and are identical to those of a random matrix with the same symmetry
(usually referred to as Wigner-Dyson statistics (WD)).
However this conjecture is not always verified. A paradigmatic example 
is the kicked rotor: 
Although classically chaotic, its spectral correlations follow Poisson
statistics and eigenfunctions are exponentially localized. 
This behavior was fully understood \cite{fishman} after mapping the kicked 
rotator problem onto an 1D disordered system (free particle in a random 
potential). It is by now well established that, in less than three dimensions,
all the eigenstates of a disordered system are exponentially localized in the 
thermodynamic limit for any amount of disorder. However, in three and higher 
dimensions, there exists a metal insulator transition (MIT), also called 
Anderson transition,  for a critical amount of disorder. At the MIT, the 
wavefunction moments ${\cal P}_q$ present anomalous scaling with respect 
to the sample size $L$, ${\cal P}_q=\int d^dr |\psi({\bf r})|^{2q}\propto 
L^{-D_q(q-1)}$, where $D_q$ is a set of exponents describing the transition. 
Wavefunctions with such a nontrivial scaling are said to be multifractal 
\cite{aoki}. Spectral fluctuations at the MIT are commonly referred to as
`critical statistics' \cite{kravtsov97}. 
Typical features include: scale invariant spectrum \cite{sko}, level repulsion 
and linear number variance \cite{chi}.
Similar properties have also been observed in disordered systems with random 
power-law hopping \cite{levitov1} (linked to classical anomalous diffusion 
\cite{ant7}) 
provided that the exponent of the hopping decay matches the dimension of the 
space. 
In the 1D case ($1/r$ decay) \cite{prbm} the relation to a MIT has been 
analytically established. For {\it non-random} power-law hopping a MIT was
recently reported\cite{sierra} even for power-law exponents larger than the 
dimensionality of the space. Unlike the random case, critical states appear
only in a certain energy window. A similar power-law behavior has also been
found in certain non-random Hamiltonians: Fermi accelerator \cite{cordero}, 
Coulomb billiard\cite{altshu}, Anisotropic Kepler problem \cite{wintgen} and 
generalized Kicked rotors \cite{bao,liu}. In all of them the potential has 
a singularity and consequently the KAM theorem does not hold. 
The classical dynamics of these systems is intermediate between chaotic and 
integrable and the classical phase space is homogeneously filled with cantori. 
The goal of this letter is to show that, in certain cases, classical chaotic 
Hamiltonians with non-analytical potentials present classical anomalous 
diffusion and quantum power-law localization of the eigenstates. The case
of logarithmic singularities, corresponding to a MIT, will be investigated
in detail.\\
{\it Classical mechanics--}We study the Hamiltonian
\be
\label{ourmodel}
{\cal H}= \frac{p^2}2 +V(q)\sum_n\delta(t -nT)
\ee
with $V(q)=\epsilon |q|^\alpha$ and $V(q)=\epsilon\log(|q|)$; $q\in [-\pi,\pi)$,
$\alpha \in [-1,1]$ and $\epsilon$ a real number. The dynamical evolution over
$T$ is dictated by the map:
$p_{n+1}=p_{n}- \frac{\partial V(q_n)}{\partial q_n}$, $q_{n+1}=q_n+Tp_{n+1}$ 
(mod$~2\pi$). 
Our aim is to show that, though obviously deterministic, the motion is accurately
modeled by an anomalous diffusion process. In particular we compute the classical
density of probability $P(p,t)$ for finding the particle at time $t$ with a 
momentum $p$. Unlike the case of normal diffusion, the information obtained from
the knowledge of a few moments of the distribution is not sufficient to fully 
characterize the classical motion. For instance, the second moment may be $\langle
r^2 \rangle \sim t$ but this by no means assures that the density of probability 
is Gaussian like \cite{klafter}.\\
\begin{figure}
\includegraphics[width=.95\columnwidth]{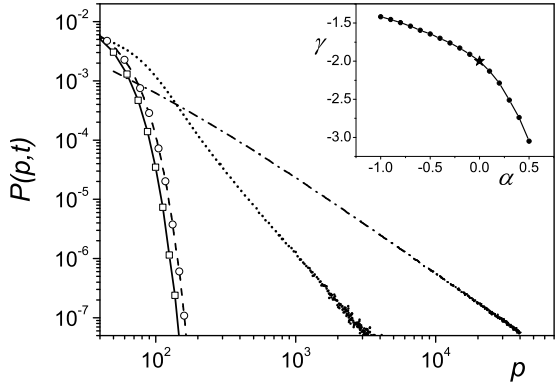}
\vspace{-.7cm}
\caption{$P(p,t)$ versus $p$ ($\log$ scale). In all cases $t=2000$ except for 
$\alpha=-0.5$ where $t=40$. For $V(q)=|q|^{\alpha}$ with $\alpha=0.9$ (solid) 
and $V(q)=(|q|+0.005)^{0.4}$ (dash) the diffusion is normal (squares and circles
are for best Gaussian fitting respectively). However for $\alpha=0.4$ (dot) and 
$\alpha=-0.5$ (dash-dot), $P(p,t)\propto t p^{\gamma(\alpha)}$. Inset shows 
$\gamma$ as a  function of $\alpha$. The star corresponds to $V(q)=\log(|q|)$ 
(see Fig.2).}
\label{figure1}
\end{figure}
\begin{figure}
\includegraphics[width=.95\columnwidth]{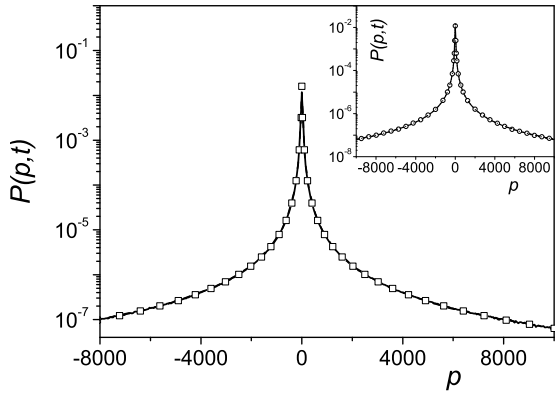}
\vspace{-.7cm}
\caption{$P(p,t=40)$ versus $p$ (semi-log scale). For $V(q)=\log(|q|)$ (solid) 
$P(p,t)$ is very well fitted by the Lorentzian distribution $P(p,t)=\frac{1}{\pi}
\frac{\nu(t)}{1+p^2\nu^2(t)}$ with $\nu(t)=2/t$ (squares). 
As shown in the inset, $P(p,t)$ (solid) is not affected by adding
a noise term $\cos(q)$ to $V(q)$ (circles).}
\label{figure2}
\end{figure}
The classical density of probability $P(p,t)$ has been evaluated by evolving 
$10^8$ initial conditions uniformly distributed in $(q,p)\in[-\pi,\pi)\times
[-\pi,\pi)$.  For $\alpha> 0.5$ (see Fig. 1), the diffusion is normal and 
$P(p,t) \sim e^{-cp^2/t}/\sqrt{t}$ ($c$ a constant). However, for 
$-0.5<\alpha < 0.5$ and $t > p$, $P(p,t) \sim t p^{\gamma(\alpha)}$ has 
power-law tails (see inset in Fig. 1 for a relation between $\gamma$ and 
$\alpha$). Such tails are considered a signature of anomalous diffusion. 
They are indeed a solution of the fractional diffusion equation, 
$
(\frac{\partial}{\partial t}- \frac{\epsilon}{2}\frac{\partial^{-\gamma-1}}
{\partial |p|^{-\gamma-1}})P(p,t)= \delta(p)\delta(t)
$
(see \cite{rev} for the definition of fractional derivative). In order to 
explicitly show that the origin of the anomalous diffusion is exclusively 
related to the classical singularity we have repeated the calculation with
the singularity-free potential $V(q)=\epsilon (|q|+b)^\alpha$, $b >0$. As 
expected, the diffusion goes back to normal as in the kicked rotor with a 
smooth potential (see Fig. 1). We have also checked that our results are 
generic, namely, they do not depend on the details of the potential but only
on the kind of singularity. Thus $P(p,t)$ is not altered by adding a smooth 
perturbation $V_{per}$ in Eq.(\ref{ourmodel}) (see inset of Fig. 2). The 
potential $V(q)=\epsilon \log(|q|)$ has been studied in detail since its
quantum counterpart (see below) resembles a disordered conductor at the MIT. 
In this case (see Fig. 2) $P(p,t)$ is accurately fitted by a Lorentzian. 
We recall this special form of $P(p,t)$ describes the dynamics of systems
with $1/f$ noise \cite{rev}. In passing we mention that in this case the 
classical phase space is dominated exclusively by cantori with no 
coexistence of chaotic and integrable regions.\\ 
{\it Quantum mechanics--} We now discuss the quantum properties of 
Eq.(\ref{ourmodel}). For the sake of clearness let us  first state our main 
results:\\
1. The eigenstates of the evolution matrix associated to the Hamiltonian 
Eq. (\ref{ourmodel}) are power-law localized with a exponent depending only 
on the singularity $\alpha$.\\
2. For $V(q)=\epsilon\log(|q|)$, the eigenstates are multifractals and the 
spectral correlations are described by critical statistics as at the MIT.\\
\noindent
The quantum dynamics of a periodically kicked system is governed by 
the quantum evolution operator $\cal U$ over a period $T$. Thus, after
a period $T$, an initial state $\psi_0$ evolves to 
$\psi(T) = {\cal U}\psi_0 = e^{\frac{-i {\hat p}^2T}{4{\bar h}}}
e^{-\frac{iV(\hat q)}{\bar h}}e^{\frac{-i {\hat p}^2 T}{4{\bar h}}}\psi_0$ 
where $\hat p$ and $\hat q$ stand for the usual momentum and position 
operator.  
Our aim is to solve the eigenvalue problem 
${\cal U}\Psi_{n}=e^{-i\kappa_n/ \hbar}\Psi_{n}$ where $\Psi_{n}$ is an 
eigenstate of $\cal U$ with eigenvalue $\kappa_n$. In order to proceed we 
first express the evolution operator in a matrix form  $\langle m| {\cal U}
| n \rangle $ in the basis of the momentum eigenstates $\{| n \rangle = 
\frac {e^{in \theta}}{\sqrt{2\pi}}\}$.
For practical calculations one option is to make the Hilbert space finite
(i.e.$~m,n=1,\ldots N$) but still keep the matrix Unitary. This can be 
achieved by imposing a periodic condition (with period $P$) to the momentum.
A large period makes the phase space effectively resemble a cylinder ($P=N$
in this paper; for more details see \cite{reviz}). 
In order to keep the kinetic term of the evolution matrix also periodic, we
take $T=2\pi M/N$ with $M$ an integer (not a divisor of $N$). $M$ is specified
to make $T$ roughly constant ($\approx0.1$ in this paper) for every $N$ used.
Based on the relation between the period in momentum $P$ and the period in 
position $Q$ ($PQ= 2\pi N\hbar$), $\hbar = 1$. The resulting evolution matrix
(for $N$ odd) then reads
\be
\label{uni}
\langle m| {\cal U}| n \rangle = \frac{1}{N}e^{-i2\pi M n^2/N}
\sum_{l}e^{i\phi(l,m,n)}
\ee
where $\phi(l,m,n)= 2\pi (l+\theta_0)(m-n)/N-iV(2\pi (l+\theta_0)/N)$,
$l = -(N-1)/2,\ldots (N-1)/2$ and $0 \le \theta_0 \le 1$; $\theta_0 $ 
is a parameter depending on the boundary conditions ($\theta_0=0$ for 
periodic boundary conditions). The eigenvalues and eigenvectors of $\cal U$
can now be computed by using standard diagonalization techniques. For 
$\theta_0=0$, parity is a good quantum number and consequently states 
with different parity must be treated separately.
\begin{figure}
\includegraphics[width=0.79\columnwidth,clip]{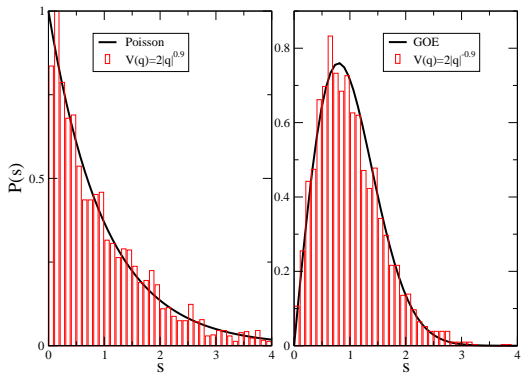}
\caption{The level spacing distribution $P(s)$ in units of the mean level 
spacing $s=(E_i-E_{i-1})/\Delta$. A transition is observed from Poisson to 
WD statistics (keeping $\epsilon$ constant) as $\alpha$ goes from negative 
to positive. In both cases $N=3100$.}
\label{figure3}
\end{figure}
\begin{figure}
\includegraphics[width=0.8\columnwidth,clip]{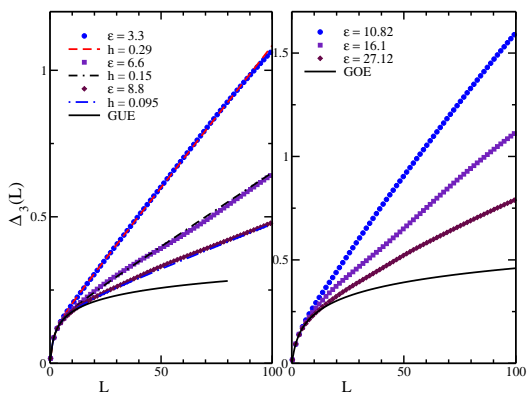}
\caption{The spectral rigidity $\Delta_3(L)$ versus $L$. Symbols are the 
numerical calculation for $V(q)=\epsilon \log (q)$. Lines are the
analytical prediction Eq.(\ref{cri}). The right (left) panel corresponds to 
the case of (broken) time reversal invariance.}
\label{figure4}
\end{figure}

We first study the spectral fluctuations of the unfolded spectrum. The analysis
of short range spectral correlations as the level spacing distribution $P(s)$ 
indicates (see Fig. 3) that for $\alpha > 1/2$ ($\alpha < -1/2$), $P(s)$ tends 
to Poisson (WD) as the volume is increased. In the intermediate region 
$-1/2<\alpha < 1/2$ long range correlators as the spectral rigidity
$\Delta_{3}(L)=\frac{2}{L^4}\int_{0}^{L}(L^3-2L^2x+x^3)\Sigma^{2}(x)dx$
where 
$\Sigma^{2}(L)=\langle L^2 \rangle - \langle L \rangle^2 = L +2\int_{0}^{L}
(L-s) R_{2}(s)ds$ is the number variance and $R_2(s)$ is the two level spectral 
correlations function, are more suitable for a qualitative analysis due to 
the possibility of performing a spectral average. We restrict ourselves to 
the potential $V(q)=\epsilon \log(q)$ related to classical $1/f$ noise. The
$R_2(s)$ associated to this type of motion has been evaluated in Ref.\cite{ant7}
for systems with broken time reversal invariance,  
\be
\label{cri}
R_2(s)=-K^2(s)=-\frac{\pi^2 h^2}{4}\frac{\sin^2(\pi s)}{\sinh^2(\pi^2 h s/2)} 
\ee
where $h \ll 1$ is related to $\epsilon$ by $h =1/\epsilon \ll 1$.
In Fig. 4 we show $\Delta_3(L)$ for different $\epsilon$'s and $N=3099$.
The time reversal invariance in Eq.(\ref{uni}) is broken by setting $T=2\pi 
\beta$ with $\beta$ an irrational number. As observed in Fig. 4 (left), 
$\Delta_3(L)$ is asymptotically linear, this feature is typical of a 
disordered conductor at the MIT. Moreover, the agreement with the analytical
prediction based on Eq. (\ref{cri}) is excellent. We remark that the value 
of $h$ best fitting the numerical result is within $5 \%$ of the analytical
estimate $h = 1/\epsilon$. We have repeated the calculation keeping the time
reversal invariance ($T=2\pi M/N$) of Eq.(\ref{cri}) (Fig. 4, right). We do
not have in this case an analytical result to compare with, but $\Delta_3(L)$
is also asymptotically linear as in the previous case.
The eigenstates $\Psi_n=\sum \psi_n(k)|k\rangle$ of the evolution operator 
$\cal U$ are also strongly affected by the classical singularity $\alpha$. 
Instead of the exponential localization typical of the standard kicked rotor,
our numerical results indicate that 
$|\psi_{n}(k)| \sim |k-k_{max}|^{-(\alpha+1)}$ decays with power-law tails 
($k_{max}$ is where $|\psi_{n}(k)|$ takes the maximum). This is in agreement
with the findings of \cite{prbm}. The eigenvector calculation was carried out 
in the cylindrical phase space (i.e. we switch off the periodic 
condition in momentum). Fig. 5 shows the results after a proper average over
all $N$ eigenstates.
Remarkably, both classical (anomalous) dynamics and quantum (power-law) 
localization are controlled by the classical singularity $\alpha$. This 
fact allows us to define a new universality class in quantum chaos 
labeled by $\alpha$. 
For $V(q)=\epsilon \log(|q|)$,  $|\psi_{n}(k)| \sim 1/k$ and, according 
to \cite{prbm}, the system is at the MIT. 
We now investigate how transport properties are affected by the 
non-analytical potential. We compute the quantum density of probability 
$P_q(k,t)=|\langle k|\phi(t)\rangle |^2$ of finding a particle with momentum 
$k\hbar$ at time $t$ for a given initial state $|\phi(0)\rangle$. We restrict
ourselves to the case $V(q)=\epsilon\log(|q|)$ in order to compare our findings
with the ones at the Anderson transition where it is reported \cite{huck} that,
in the limit $k\ll t \gg 1$, $P_q(k,t)\sim t^{-D_2}k^{D_2-1}$ where $D_2$ is 
the multifractal dimension defined previously. As expected, we have observed 
a similar power-law behavior in our model (see Fig. 6). The numerical value of
$D_2$ increases as $\epsilon$ does and it is within $10\%$  of the one obtained
by the inverse participation ratio (not shown). 
These results show that quantum mechanically the diffusion is also anomalous but, 
unlike the classical case, the decay of $P_q(k,t)$ depends not only on the classical 
singularity $\alpha$ but also on the coupling constant $\epsilon$. The 
overall effect of the quantum corrections is also to suppress the classical
(anomalous) diffusion at a rate that increases as $\epsilon$ gets smaller.
To summarize, quantum 1+1D systems with classical log singularities share
all the properties of a disordered conductor at the MIT: Multifractality, 
critical statistics and quantum anomalous diffusion. 
Analytical results can in principle be 
obtained by mapping Eq.(\ref{ourmodel}) onto an
1D Anderson model. This method was introduced in \cite{fishman} for the case
of a kicked rotor with a smooth potential.
\begin{figure}
\includegraphics[width=.95\columnwidth]{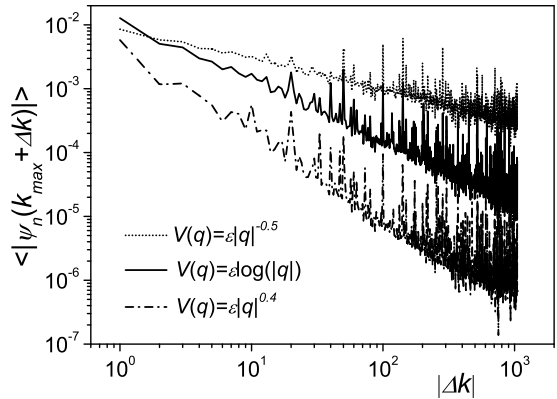}
\vspace{-.7cm}
\caption{Averaged modulus of the eigenstates of the evolution operator for
 $N=2100$. We set $\epsilon=0.01$ to a small value in order to avoid the 
 finite size effects. The best fitting power exponents $-0.49$, $-0.98$ and
 $-1.36$ correspond to $\alpha=-0.5,0 (\log),0.4$ respectively are very close
 to the analytical prediction $-(1+\alpha)$ \cite{prbm}.}
\label{figure5}
\end{figure}
\begin{figure}
\includegraphics[width=.95\columnwidth]{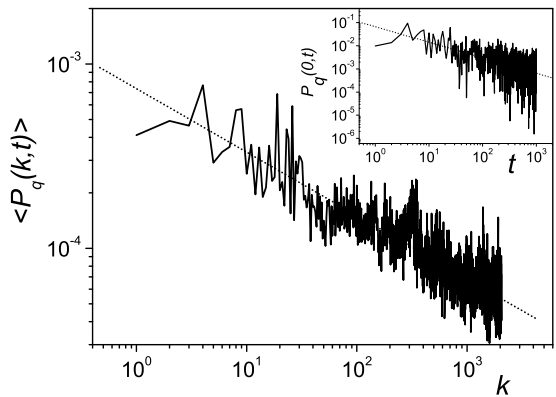}
\vspace{-.7cm}
\caption{Time averaged $P_q(k,t)$ over $4950 \le t \le 5050$ and $P_q(k=0,t)$ 
 versus $t$ (inset) for $V(q)=10\log(|q|$). The best fitting (dotted line)
 is of a slope $-0.34$ and $-0.67$ (inset) respectively, which correspond to
 a $D_2 \approx 0.67$.}
\label{figure6}
\end{figure}
We do not repeat here the details of the calculation but just state how the 
1D Anderson model is modified by the non-analytical potential. It turns out 
that classical non-analyticity induces long-range disorder in the associated
1D Anderson model, $T_mu_m+\sum_r U_r u_{m+r}=E u_m$ where 
$U_r \sim A_r/r^{1+\alpha}$ is the Fourier coefficient of $-\tan V(q)/2$, 
$T_m=\theta_0/2 -T m^2/2$ and $E=-U_0$. For random Gaussian $A_r$ this model
corresponds to the one studied in \cite{prbm} which is solved by using the 
supersymmetry method. For the special case of a constant non-random $A_r$ 
(closer to our case) it was found that \cite{sierra} some of the eigenvectors
remain critical even for $1 <\alpha$. This suggests that the results reported
above for $\alpha=0$ (log potential) may be extended to other $\alpha$'s 
though clearly more work is needed to further clarify this issue.  In 
conclusion, we have shown that classical singularities in chaotic Hamiltonians
may lead to classical anomalous diffusion and quantum power-law localization. 
Both quantum and classical features are governed by the classical singularity.
We thus suggest a new universality class in quantum chaos labeled by the type 
of classical singularity. In the case of logarithmic singularities the 
classical dynamics  presents $1/f$ noise. Quantum mechanically the system 
posses all the features of a MIT: multifractal wavefunctions, critical 
statistics and quantum anomalous diffusion.\\
AMG thanks Baowen Li's group at the National University of Singapore for 
his warm hospitality and financial support. 
We also thank Baowen Li for illuminating discussions.
AMG is supported by a 
postdoctoral fellowship from the Spanish MECD. JW is 
supported by DSTA Singapore under Project Agreement POD0001821.

\end{document}